\documentclass[twocolumn]{aastex62}
\pdfoutput=1 
\usepackage{amsmath,amstext}
\usepackage[T1]{fontenc}
\usepackage[figure,figure*]{hypcap}
\usepackage{graphicx}
\usepackage{subfigure}
\usepackage{tabularx}
\usepackage{needspace,enumitem}

\newcommand{\sw}{Sw\,1644+57}

\shorttitle{Swift J1644+57 Update}
\shortauthors{Cendes et al.}

\begin{document}

\title{Radio Monitoring of the Tidal Disruption Event Swift J164449.3+573451. IV. Continued Fading and Non-Relativistic Expansion}

\correspondingauthor{Yvette Cendes}
\email{yvette.cendes@cfa.harvard.edu}

\author{Y. Cendes}
\affiliation{Center for Astrophysics | Harvard \& Smithsonian,
Cambridge, MA 02138, USA}

\author{T. Eftekhari}
\affiliation{Center for Astrophysics | Harvard \& Smithsonian,
Cambridge, MA 02138, USA}

\author{E. Berger}
\affiliation{Center for Astrophysics | Harvard \& Smithsonian,
Cambridge, MA 02138, USA}

\author{E. Polisensky}
\affil{Remote Sensing Division, Naval Research Laboratory, Washington, DC 20375, USA}

\begin{abstract}

We present continued radio and X-ray observations of the previously relativistic tidal disruption event (TDE) Swift J164449.3+573451 (\sw) extending to about 9.4 years post disruption, as part of ongoing campaigns with the Jansky Very Large Array (VLA) and the \textit{Chandra} X-ray observatory.  We find that the X-ray emission has faded below detectable levels, with an upper limit of $\lesssim 3.5\times 10^{-15}$ erg cm$^{-2}$ s$^{-1}$ in a 100 ks observation, while the radio emission continues to be detected and steadily fade.  Both are consistent with forward shock emission from a non-relativistic outflow, although we find that the radio spectral energy distribution is better fit at these late times with an electron power law index of $p\approx 3$ (as opposed to $p\approx 2.5$ at earlier times).  With the revised spectral index we find $\epsilon_B\approx 0.01$ using the radio and X-ray data, and a density of $\approx 0.04$ cm$^{3}$ at a radius of $R\approx 0.65$ pc ($R_{\rm sch}\approx 2\times 10^6$ R$_\odot$) from the black hole. The energy scale of the blastwave is $\approx 10^{52}$ erg. We also report detections of \sw\ at 3 GHz from the first two epochs of the VLA Sky Survey (VLASS), and find that $\sim 10^2$ off-axis \sw-like events to $z\sim 0.5$ may be present in the VLASS data.  Finally, we find that \sw\ itself will remain detectable for decades at radio frequencies, although observations at sub-GHz frequencies will become increasingly important to characterize its dynamical evolution.

\end{abstract}


\section{Introduction} 
\label{sec:intro}

A tidal disruption event (TDE) occurs when a star wanders sufficiently close to a supermassive black hole (SMBH) to be torn apart by tidal forces.  In recent years, the number of observed TDEs has increased dramatically, primarily thanks to wide-field optical time-domain surveys\footnote{See http://tde.space}.  So far, only about 10 TDEs (about 10\% of the known sample) have been detected in the radio band during dedicated follow-up observations \citep[see Table 1; ][]{Alexander2020}.  In these cases, the radio emission points to non-relativistic outflows with an energy scale of $\sim 10^{37}-10^{39}$ erg s$^{-1}$. 

The TDE \sw\ remains unusual in this context. It was first detected as a $\gamma$-ray/X-ray event on 2011 March 28 by the {\it Swift} Burst Alert Telescope (BAT), although further analysis showed discernible emission as early as 2011 March 25 \citep{Burrows2011}. Multi-wavelength observations in the radio, millimeter, near-IR, and optical were obtained shortly after discovery, and along with the high energy emission pointed to the launch of a relativistic jet \citep{Bloom2011,Burrows2011,Zauderer2011,p1}.  Subsequently, two additional potential relativistic TDEs have been discovered \citep{Cenko2012,Brown2015}, although \sw\ remains by far the best-studied case. 

Continued monitoring of \sw\ revealed a sharp decline by about 2 orders of magnitude in its X-ray emission $\approx 500$ days post disruption, corresponding to the relativistic jet shutting off \citep{p2}. The subsequent X-ray emission has instead been shown to be an extension of the spectral energy distribution (SED) from the radio band, produced by the interaction of the blastwave with the ambient medium. The early radio/mm emission, and subsequent addition of X-ray information, have allowed us to track the velocity and energy scale of the blastwave, as well as the density profile of the circumnuclear medium \citep{p1,p2,p3}.

Here, we present continued X-ray and radio observations of \sw\ on a timescale of about 2390 to 3430 days post disruption.  We use these most recent observations to  continue tracking the blastwave energy and circumnuclear medium density profile.  In Section~\ref{sec:obs}, we present our new observations.  In Section~\ref{sec:modeling} we model the radio data, in conjunction with the X-ray data determine the location of the cooling frequency, and carry out an equipartition analysis to determine the physical properties of the outflow and ambient medium.  In section~\ref{sec:discussion} we discuss our findings in the context of the on-going evolution of \sw, and the implications of its detection in VLASS.  We summarize our  conclusions in Section~\ref{sec:conclusions}.

\section{Observations}
\label{sec:obs}

Our previous work presented radio and mm observations of \sw\ extending to about 1895 d post disruption, and X-ray observations extending to about 2020 d post disruption \citep{Zauderer2011,p1,p2,p3}.  Here we present continued radio and X-ray observations.

\subsection{Radio Observations}

We obtained radio observations of \sw\ with the Karl G.~Jansky Very Large Array (VLA) spanning from L- to K-band ($1-26.5$ GHz; see Table~\ref{tab:obs}; Program ID: 17B-198; PI: Eftekhari).  All times are measured relative to 2011 March 25, corresponding to the time of first discernible $\gamma$-ray emission.  We processed the data using the Common Astronomy Software Application package (CASA; \citealt{McMullin2007}) accessed through the python-based \texttt{pwkit} package\footnote{https://github.com/pkgw/pwkit} \citep{Williams2017}.  We performed bandpass and flux density calibration using 3C286 for all observations and frequencies and J1638+5720 for the phase calibrator.  We obtained all flux densities and uncertainties via the \texttt{imtool} program in \texttt{pwkit}.

In addition to our dedicated observations, we examined observations from the Very Large Array Sky Survey \citep[VLASS; ][]{Lacy2020} at a frequency of 3 GHz (S-band) and report a detection of $1.2\pm 0.2$ mJy in the ``quick look'' images from 2017 October 7 (Day 2388), and a detection of $0.3\pm 0.2$ mJy on 2020 August 4 (Day 3434; see Table~\ref{tab:obs}).  Additionally, we identified data collected by the VLA Low-band Ionosphere and Transient Experiment \citep[VLITE; ][]{Clarke2016} during our dedicated observation on  2018 January 18 (Day 2493) and report a $3\sigma$ upper limit at 350 MHz of $\lesssim 2.5$ mJy (Table~\ref{tab:obs}).

\begin{deluxetable}{lccc}
\tablecolumns{4}
\tablecaption{VLA Radio Observations}
\tablehead{
Date       &
$\delta t$ &  
Frequency  & 
Flux Density \\
(UTC) & 
(d)   & 
(GHz) & 
(mJy)
}
\startdata
2017 Oct 7 & 2388$^*$ & 3.0 & $1.2\pm 0.2$ \\\hline
2018 Jan 20 & 2493 & 0.35 & $<2.5$ \\
&     & 1.5 & $1.04\pm 0.06$ \\
&     & 2.6 & $1.39\pm 0.03$ \\
&     & 3.4 & $1.29\pm 0.03$ \\
&     & 6.0 & $0.84\pm 0.01$ \\
&     & 9.8 & $0.53\pm 0.01$ \\
&     & 14.7 & $0.34\pm 0.01$ \\
&     & 21.8 & $0.22\pm 0.01$ \\\hline
2018 Nov 17  & 2795 & 1.5 & $<0.3$ \\
&     & 3.0  & $0.97\pm 0.03$ \\
&     & 6.0  & $0.60\pm 0.01$ \\
&     & 9.8  & $0.40\pm 0.02$ \\
&     & 14.7 & $0.25\pm 0.01$ \\
&     & 21.8 & $0.16\pm 0.03$ \\\hline
2020 Aug 4 & 3434$^*$& 3.0& $0.3\pm 0.2$\\
\enddata
\tablecomments{New VLA observations of \sw, with upper limits quoted as  $3\sigma$.\\
$^*$ Flux density measurement from the VLA Sky Survey.}
\label{tab:obs}
\end{deluxetable}

\subsection{X-Ray Observations}

\begin{deluxetable*}{cccccc}
\tablecolumns{5}
\tablecaption{\textit{Chandra} X-ray Observations}
\tablehead{
	\colhead{Date} &
	\colhead{$\delta t$} &
	\colhead{Exposure Time} &
	\colhead{Net Counts} &
	\colhead{Net Count Rate} &
	\colhead{Flux ($0.3 - 10$ keV)} \\
	\colhead{(UTC)} &
	\colhead{(d)} &
	\colhead{(ks)} &
	\colhead{} &
	\colhead{(cts/s)} &
	\colhead{(ergs/cm$^2$/s)}	
}
\startdata
2015 Feb 20 & 1425 & 27.8&12.9 $\pm$ 3.6 &(4.6 $\pm$ 1.3) $\times10^{-4}$&  (1.5 $\pm 0.4) \times 10^{-14}$\\
2015 April 9 & 1473 & 18.7&3.9 $\pm$ 2.0 & (2.1 $\pm$ 1.0) $\times10^{-4}$& (7.3 $\pm 3.5) \times 10^{-15}$\\
2015 Aug 25 & 1611 & 19.7&0 &< 4.1 $\times10^{-4}$ & < 1.4 $\times 10^{-14}$ \\ 
2016 Feb 1 & 1771 & 26.7&6.4 $\pm$ 2.4& (2.4 $\pm$ 1.0) $\times10^{-4}$&  (8.5 $\pm 4.0) \times 10^{-15}$\\
2016 April 6 & 1836 & 24.6&5.8 $\pm$ 2.4 & (2.3 $\pm$ 1.0) $\times10^{-4}$& (8.4 $\pm 4.0) \times 10^{-15}$\\
2016 19 June & 1910 & 24.6&4.8 $\pm$ 2.1 & (1.9 $\pm$ 0.9) $\times10^{-4}$& (6.7 $\pm 3.2) \times 10^{-15}$\\
2016 Oct 3 & 2019 & 24.6&2.6 $\pm$ 1.5 & (1.1 $\pm$ 0.6) $\times 10^{-4}$& (3.9 $\pm 2.1) \times 10^{-15}$\\
\hline
2018 Feb $5-11$ & 2516 & 47.2 & 2.3 $\pm$ 2.7 & < 1.3 $\times10^{-4}$ & < 5.3 $\times 10^{-15}$ \\
2018 Aug 6 & 2690 & 50.9& 2.0 $\pm$ 3.0 &< 1.2 $\times10^{-4}$  & < 4.8 $\times 10^{-15}$ \\ 
\enddata
\tablecomments{Observed X-ray counts and source significance (1$\sigma$) for \sw. Unabsorbed X-ray fluxes are calculated assuming an absorbed power law model with $\Gamma_X = 2.5$ (that is, assuming $p=3$), $N_{\rm H,int} = 2 \times 10^{22} \ \rm cm^{-2}$, and $N_{\rm H,MW} = 1.7 \times 10^{20} \ \rm cm^{-2}$. Upper limits correspond to 3$\sigma$ limits. Data for $\delta t = 1425,1473$ are from \citet{Levan2016} (PI: Levan), and all other data before Day 2513 is from \citep{p3}, although the fluxes reported have been recalculated for the power law model described above.}
\label{tab:x-ray}
\end{deluxetable*}

The {\it Chandra} X-ray observations were obtained on 2018 February $5-11$ and August 6 (Day 2516: 47.2 ks; Day 2690: 50.9 ks; Program IDs: 19700497 and 19700757; PI: Eftekhari) using the Advanced CCD Imaging Spectrometer (ACIS-S).  We analyzed the observations using standard ACIS data filtering, using $\texttt{CALDB}$ within the $\texttt{CIAO}$ software package (v4.12).  For source detection, we performed a targeted extraction using an aperture size of $1.5''$ and an annular background region with inner and outer radii of $2''$ and $6''$, respectively, obtaining an individual net count rate and source significance for each observation.

\begin{figure}
    \includegraphics[width=.95\columnwidth]{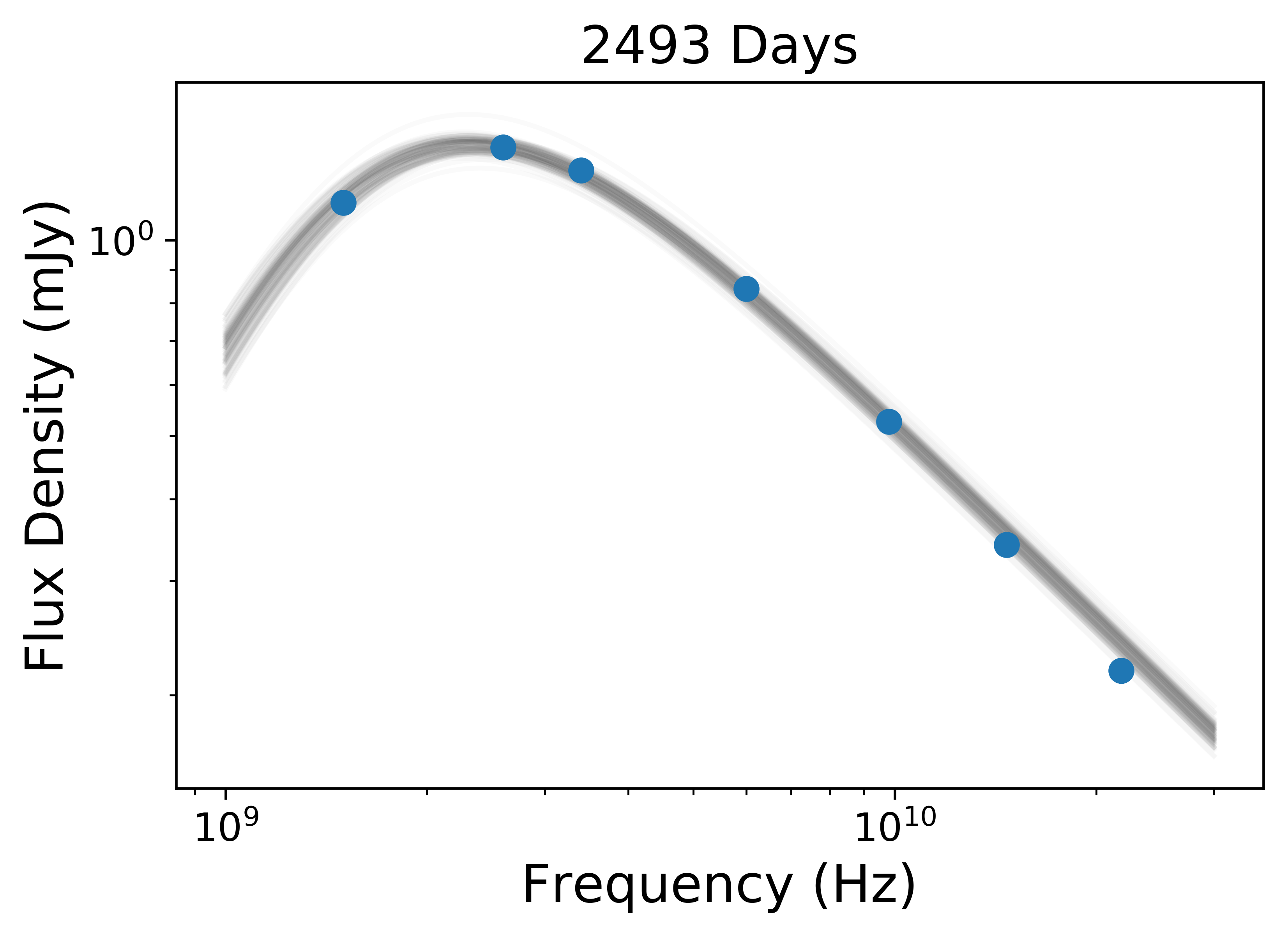}
    \label{fig:2493}
    \includegraphics[width=.95\columnwidth]{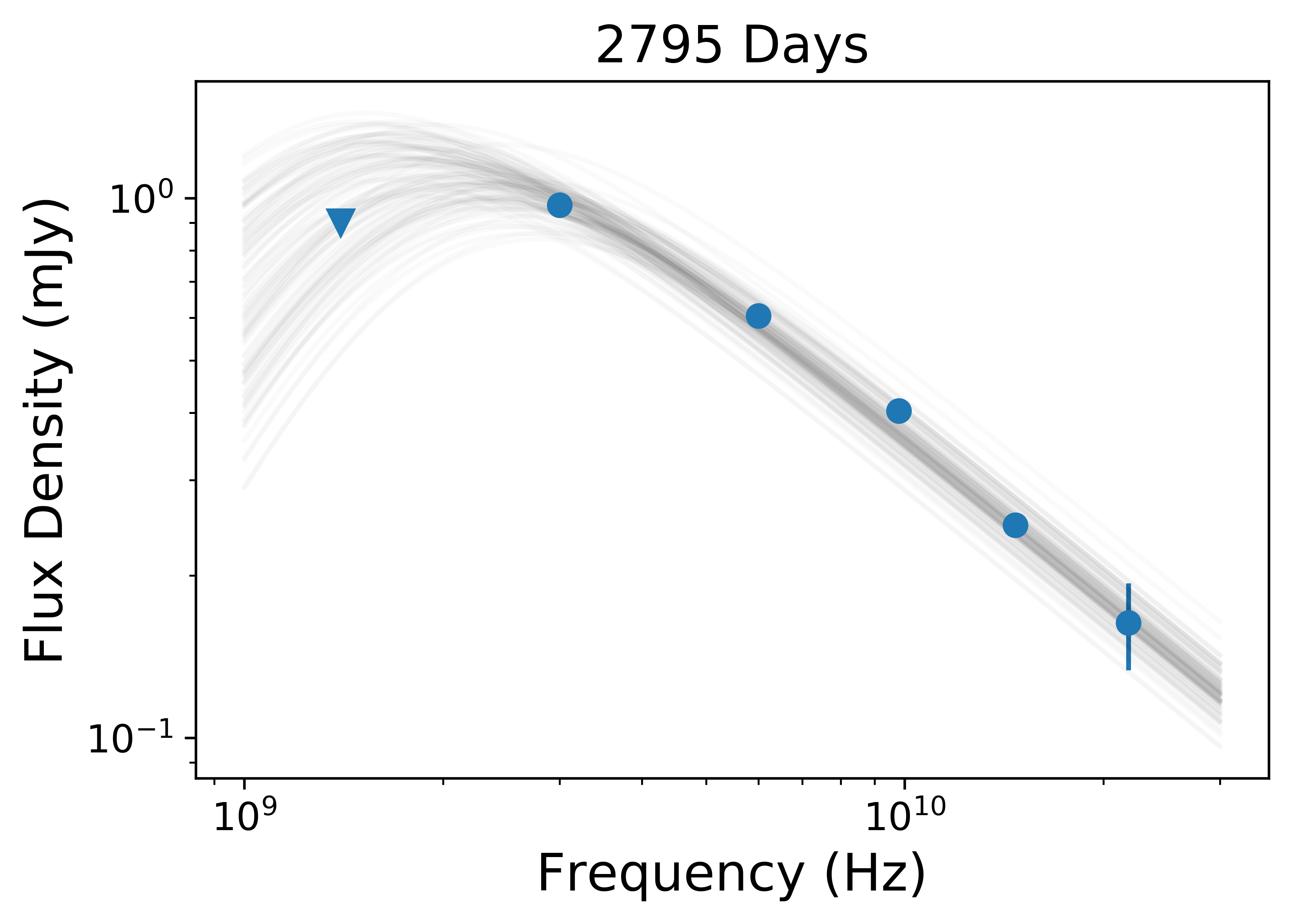}
    \label{fig:2795}
    \caption{The radio spectral energy distributions at Days 2493 and 2795.  The circles represent detections, and the triangle indicates a $3\sigma$ upper limit.  The black lines are representative fits using MCMC with $\nu_{a}, \nu_{m}$, and $F_{\nu}$ as free parameters, and using a fixed value of $p=3$.}
\end{figure}

Both of our observations yielded non-detections (see Table~\ref{tab:x-ray}).  We calculated the unabsorbed fluxes ($0.3-10$ keV) with $\texttt{PIMMS}$ (v4.10), based on the assumption of forward shock emission where $\nu_X>\nu_c$ and $p=3$ (see below), leading to a power law index of $\Gamma_X=2.5$.  \citet{p3} provided flux values for Days 1425 through 2019 assuming $p=2.5$, and we therefore recalculate these fluxes with $\Gamma_X=2.5$ for consistency.  We assumed an absorbed power law model with an intrinsic absorption column $N_{\rm H,int} = 2 \times 10^{22} \ \rm cm^{-2}$ \citep{Mangano2016}, and a Galactic absorption of $N_{\rm H,MW} = 1.7 \times 10^{20} \ \rm cm^{-2}$ \citep{Willingale2013}.

\section{Modeling and Analysis}
\label{sec:modeling}

\subsection{Spectral Energy Distribution}
\label{sec:sed}

As in our previous analyses of the radio emission from \sw, we model the spectral energy distribution (SED) as self-absorbed synchrotron emission in order to determine the physical properties of the outflow and surrounding medium as a function of time.  The SEDs at Days 2493 and 2795 are shown in Figure~\ref{fig:2795}. The SEDs are defined by a peak flux density, $F_{\nu,p}$, at a peak frequency, $\nu_{p}$, and by a power law slope related to the electron energy distribution, $p$.  We previously found that at early times $\nu_p$ is defined by $\nu_{m}$, the frequency associated with the minimum electron Lorentz factor of the electron distribution, $\gamma_m$, while at later times it is defined by the self-absorption frequency, $\nu_a$ \citep{p1,p2,p3}.

\begin{figure}
    \includegraphics[width=\columnwidth]{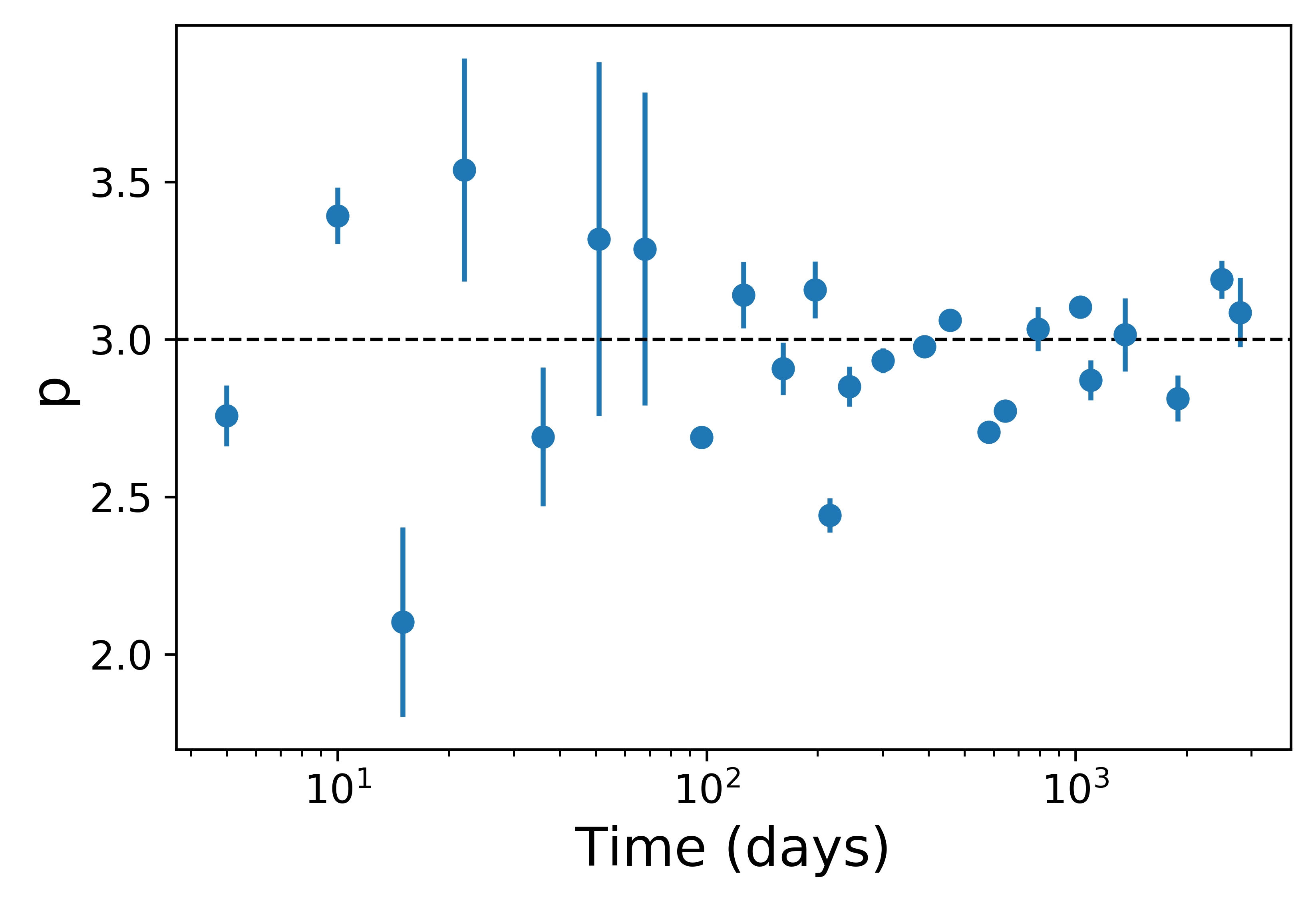}
    \caption{ Best-fit values for the the electron power law index, $p$, as a function of time from our MCMC modeling of the individual radio/mm SEDs.  The black dotted line marks the weighted mean value of $p\approx 3.0$ at $\gtrsim 100$ d, which is the value used in this work.}
    \label{fig:vary-p}
\end{figure}

Following \citet{p3}, we fit the SED with a weighted model that accounts for both possible versions of the peak frequency, using the models developed by \citet{Granot2002} for synchrotron emission from GRB afterglows.  When $\nu_{a}\ll\nu_{m}$ we have:
\begin{multline}
\indent F_1 \equiv F_\nu (\nu_a) \Bigg[ \Big(\frac{\nu}{\nu_a}\Big)^{-s_1\beta_1}+\Big(\frac{\nu}{\nu_a}\Big)^{-s_1\beta_2}\Bigg] \ \times 
\\ 
\Bigg[1+\Big(\frac{\nu}{\nu_m}\Big)^{-s_2 (\beta_2-\beta_3)}\Bigg]^{-1/s_2},
\label{eq:first-spec}
\end{multline}
where $\beta_n$ describes the spectral slopes above and below a break, and $s_1$ and $s_2$ are smoothing parameters that describe the shape of the spectrum across each break. Here, $\beta_1 = 2$, $\beta_2=1/3$, and $\beta_3 = (1-p)/2$. When $\nu_{m}\ll\nu_{a}$, the spectrum is instead given by:
\begin{multline}
\indent F_2 \equiv F_\nu (\nu_m) \Bigg[ \Big(\frac{\nu}{\nu_m}\Big)^2 e^{-s (\frac{\nu}{\nu_m})^{2/3}}+\Big(\frac{\nu}{\nu_m}\Big)^{5/2}\Bigg] \ 
\times \\
\Bigg[1+\Big(\frac{\nu}{\nu_a}\Big)^{-s_2 (\beta_2-\beta_3)}\Bigg]^{-1/s_2},
\label{eq:second-spec}
\end{multline}
where $\beta_2 = 5/2$ and $\beta_3 = (1-p)/2$.  To obtain a smooth transition between the two regimes, we employ a weighted spectrum (Equation 3 in \citealt{p3}):
\begin{equation}
F = \frac{w_1F_1 + w_2F_2}{w_1 + w_2},
\label{eq:weighted}
\end{equation}
where $w_1 = (\nu_m/\nu_a)^2$ and $w_2 = (\nu_a/\nu_m)^2$. 

We determine the best fit using the Python Markov Chain Monte Carlo (MCMC) module \texttt{emcee} \citep{Foreman-Mackey2013}, assuming a Gaussian likelihood for the parameters $F_{\nu}$, $\nu_{a}$, and $\nu_{m}$.  We also included a parameter that accounted for systematic uncertainty for underestimates on the individual data points.  Our priors were based on the assumption that all these parameters decrease over time, as seen in \citet{p3}, and thus the value from the previous epoch was used as the upper limit for each respective parameter.  The posterior distributions were sampled using 100 Markov chains, which were run for 2,000 steps, discarding the first 1,000 steps to ensure the samples have sufficiently converged.

In \citet{p1,p2,p3}, a value of $p=2.5$ was adopted at earlier epochs, but here we fit for $p$ as a free parameter.  For this purpose we refit the entire radio data set and show the results for $p$ in Figure~\ref{fig:vary-p}. At early time ($\lesssim 100$ days) the value of $p$ exhibits large variations and uncertainty.  This is due to the high value of $\nu_p$, and hence the paucity of data beyond the peak, as well as to interstellar scintillation, which leads to fluctuations in the observed flux density and hence the model fit \citep{p1}. At later times we still find some scatter in $p$ (primarily due to residual scintillation effects) but overall the weighted average after Day 100 is $p=2.96\pm 0.06$.  We therefore adopt $p=3$ in the remainder of our analysis, and refit all of the radio data beyond Day 700 with this value.  We note that we explicitly checked the change in $p$ by slicing the data to calculate the weighted average of $p$ over time, but concluded the change in $p$ is not a physical occurrence but instead due to more precise measurements at later times.

\begin{figure}
    \includegraphics[width=1\columnwidth]{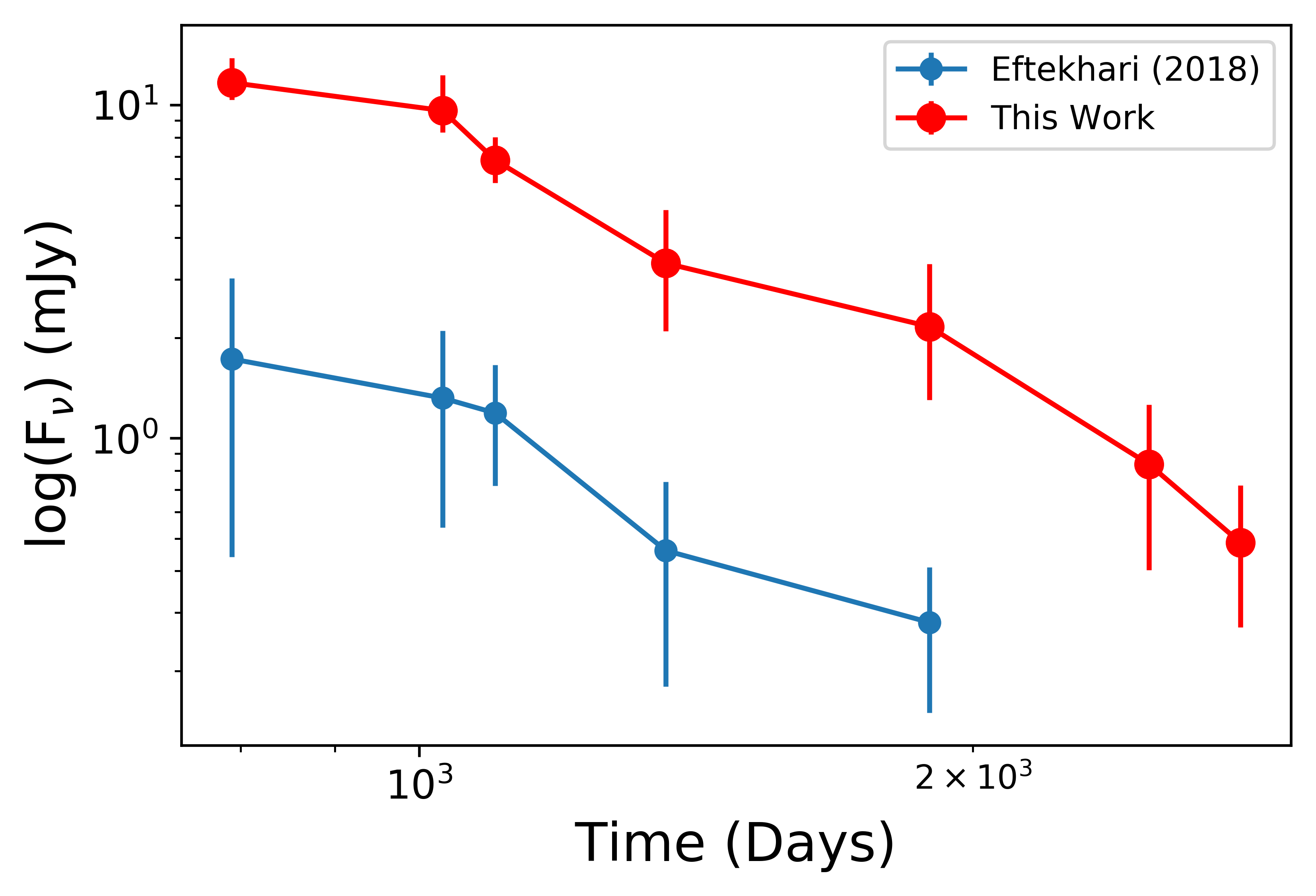}
    \label{fig:F_nu}
    \includegraphics[width=1\columnwidth]{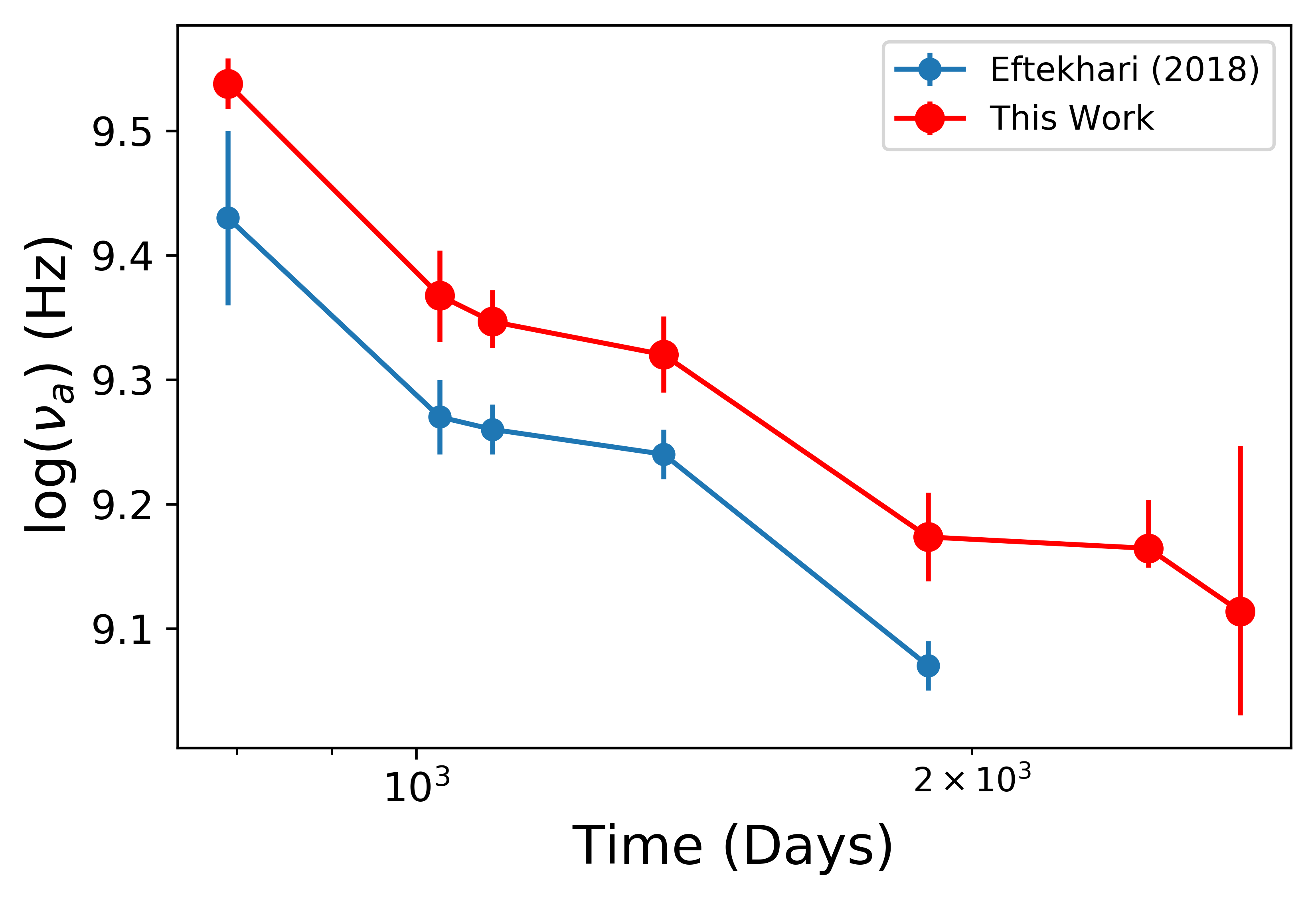}
    \label{fig:nu-a}
    \includegraphics[width=1\columnwidth]{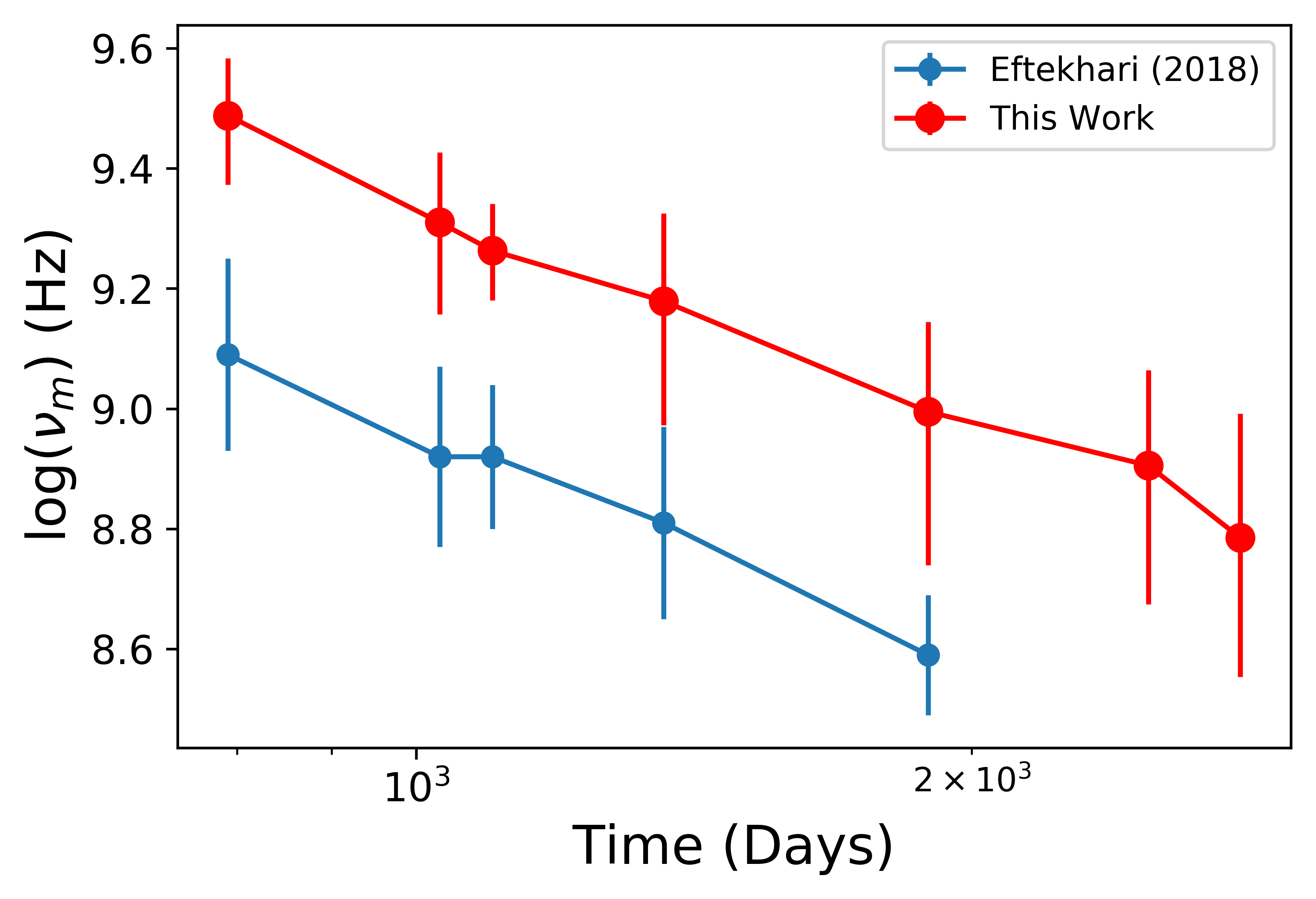}
    \caption{Time evolution of the parameters defining the radio SED from our epoch-by-epoch modeling here (red circles): $F_{\nu,p}$ ({\it Top}), $\nu_a$ ({\it Middle}), and $\nu_m$ ({\it Bottom}).  We also show the results for the same parameters from  \citet[][blue]{p3}.  The two sets of values exhibit the same temporal trend, but differ in normalization due to the use of $p=3$ in this work compared to $p=2.5$ in \citet{p3}.}
    \label{fig:nu-m}
\end{figure}

Following fitting with Equation~\ref{eq:weighted}, the resulting values for $F_{\nu}$, $\nu_a$, and $\nu_m$ are listed in Table~\ref{tab:SED-fits} and plotted in Figure~\ref{fig:nu-a}.  We note explicitly that $F_{\nu}$ is the fitted flux density, which, due to the smoothing of the SED, is larger than the actual peak flux density, $F_{p}$, used in \S\ \ref{sec:equi}.  As in \citet{p3} we find an overall steady decline in $\nu_a$ and $\nu_m$, with a corresponding decline in $F_{\nu,p}$.  However, due to the difference in the adopted values of $p$ we find systematic offsets of about a factor of 2.5 times higher for $\nu_a$, a factor of 1.3 times higher for $\nu_m$, and a factor of 7 times higher for $F_{\nu,p}$.

\begin{deluxetable*}{ccccc}
\label{tab:SED-fits}
\tablecolumns{5}
\tablecaption{Spectral Energy Distribution Model Parameters}
\tablehead{
	\colhead{$\delta t$} &
	\colhead{$F_{\nu,p}$} &
	\colhead{log($\nu_a$)} &
	\colhead{log($\nu_m$)} &
	\colhead{log($\nu_c$)} \\
	\colhead{[d]} &
	\colhead{[mJy]} &
	\colhead{[Hz]} &
	\colhead{[Hz]} &
	\colhead{[Hz]} 
}
 \startdata
 791&$11.68\substack{+2.17 \\ -1.30}$&$9.54\substack{+0.02 \\ -0.02}$&$9.49\substack{+0.10 \\ -0.11}$&$13.29\substack{+0.02 \\ -0.02}$ \\
 1030&$9.82\substack{+2.67 \\ -1.34}$&$9.37\substack{+0.04 \\ -0.04}$&$9.31\substack{+0.12 \\ -0.15}$&$13.44\substack{+0.04 \\ -0.04}$ \\
 1100&$6.83\substack{+1.18 \\ -0.99}$&$9.35\substack{+0.02 \\ -0.02}$&$9.26\substack{+0.08 \\ -0.08}$&$13.44\substack{+0.03 \\ -0.03}$ \\
 1362&$3.35\substack{+1.49 \\ -1.25}$&$9.32\substack{+0.03 \\ -0.03}$&$9.18\substack{+0.15 \\ -0.20}$&$13.42\substack{+0.05 \\ -0.05}$ \\
 1894&$2.16\substack{+1.18 \\ -0.86}$&$9.17\substack{+0.04 \\ -0.03}$&$9.00\substack{+0.15 \\ -0.25}$&$13.54\substack{+0.05 \\ -0.05}$ \\	
 2493&$0.84\substack{+0.42 \\ -0.43}$&$9.16\substack{+0.04 \\ -0.02}$&$8.91\substack{+0.15 \\ -0.23}$&$13.51\substack{+0.06 \\ -0.04}$ \\	
 2795&$0.49\substack{+0.23 \\ -0.22}$&$9.11\substack{+0.13 \\ -0.08}$&$8.79\substack{+0.20 \\ -0.23}$&$13.52\substack{+0.1 \\ -0.15}$ \\	
 \enddata
 \tablecomments{The model is described in \S\ref{sec:modeling} and in \citet{Granot2002}.  We note that the values for $\nu_c$ are inferred from the equipartition calculation.}
 \end{deluxetable*}

\subsection{Equipartition Analysis}
\label{sec:equi}

Using the inferred frequency and flux density parameters from the SED fitting in \S\ref{sec:sed}, we now infer the physical properties of the outflow using an equipartition analysis.  In \S\ref{sec:cooling} we use the X-ray data to determine any deviation from equipartition, as found in our previous work \citep{p3}.  In our analysis we follow the prescription of \citet{Duran2013}. As already shown in \citet{p3}, the outflow at late times is non-relativistic, leading to the following expressions for the equipartition radius and kinetic energy \citep[see Equations 21 and 25 in ][]{Duran2013}:
\begin{multline} 
R_{\rm eq}\approx (1.7\times10^{17} \textrm{cm})\times F_{p,\rm mJy}^{8/17} d_{L,28}^{16/17} \nu_{p,10}^{-1} (1+z)^{-25/17}
\\
\times f_A^{-7/17} f_V^{-1/17} (\nu_m/\nu_a)^{-1/34} 4^{1/17} \epsilon^{1/17}
\label{eq:rad}
\end{multline}
\begin{multline} 
E_{\rm eq} \approx (2.5\times10^{49} \textrm{cm})\times F_{p,\rm mJy}^{20/17} d_{L,28}^{40/17} \nu_{p,10}^{-1} (1+z)^{-37/17}
\\
\times f_A^{-9/17} f_V^{6/17} (\nu_m/\nu_a)^{-11/34} 4^{11/17}
\\ \times [(11/17)\epsilon^{-6/17}+(6/17)\epsilon^{11/17}],
\label{eq:energy}
 \end{multline}
where $d_L=1955$ Mpc is the luminosity distance, $z=0.354$ is the redshift, and $f_A$ and $f_V$ are the area and volume filling factors, respectively, where we use $f_A = \theta_{j}^2$ with $\theta_{j}=0.1$, and $f_V= \frac{4}{3}$ \citep{p1,p2,p3}. The factors of $4^{1/17}$ and $4^{11/17}$ for the radius and energy, respectively, arise from corrections to the isotropic number of radiating electrons ($N_{e,\rm iso}$) in the non-relativistic case.  Similarly, since at late times $\nu_m<\nu_a$, we include a correction factor of $(\nu_a/\nu_m)^{(p-2)/2}$ in $N_{e,\rm iso}$, which leads to factors of $(\nu_a/\nu_m)^{(2-p)/34}$ in $R_{\rm eq}$ and $(\nu_a/\nu_m)^{11(2-p)/34}$ in $E_{\rm eq}$.  We further assume that the fraction of post-shock energy in relativistic electrons is $\epsilon_e=0.1$, which leads to correction factors of $\xi^{1/17}$ and $\xi^{11/17}$ in $R_{\rm eq}$ and $E_{\rm eq}$, respectively, where $\xi = 1 + \epsilon_e^{-1} \approx 11$.  Finally, we parameterize a deviation from equipartition with a correction factor $\epsilon = (11/6)(\epsilon_B/\epsilon_e)$, where $\epsilon_B$ is the fraction of post-shock energy in magnetic fields.

The magnetic field and number of radiating electrons in the non-relativistic case are given by \citep[see Equations 16 and 15 in ][]{Duran2013}:
\begin{multline} 
B \approx (1.3\times10^{-2}\, \textrm{G})\times  F_{p,\rm mJy}^{-2} d_{L,28}^{-4} (1+z)^{7} f_A^{2} R_{\rm eq,17}^{4} \nu_{p,10}^5
\label{eq:mag}
\end{multline}
\begin{multline} 
N_{e} \approx (1\times10^{54})\times F_{p,\rm mJy}^{3} d_{L,28}^{6} (1+z)^{-8} f_A^{-2} R_{\rm eq,17}^{-4} \nu_{p,10}^5 \\ \times (\nu_{a}/\nu_{m})^{(p-2)/2}.
\label{eq:dens}
\end{multline}
We calculate the number density of radiating electrons as $n_{\rm ext} = 4N_e/V$, where $V = 4/3 \pi R^3$ is the volume of the emission region. 

\begin{figure}
\begin{center}
    \includegraphics[width=.75\columnwidth]{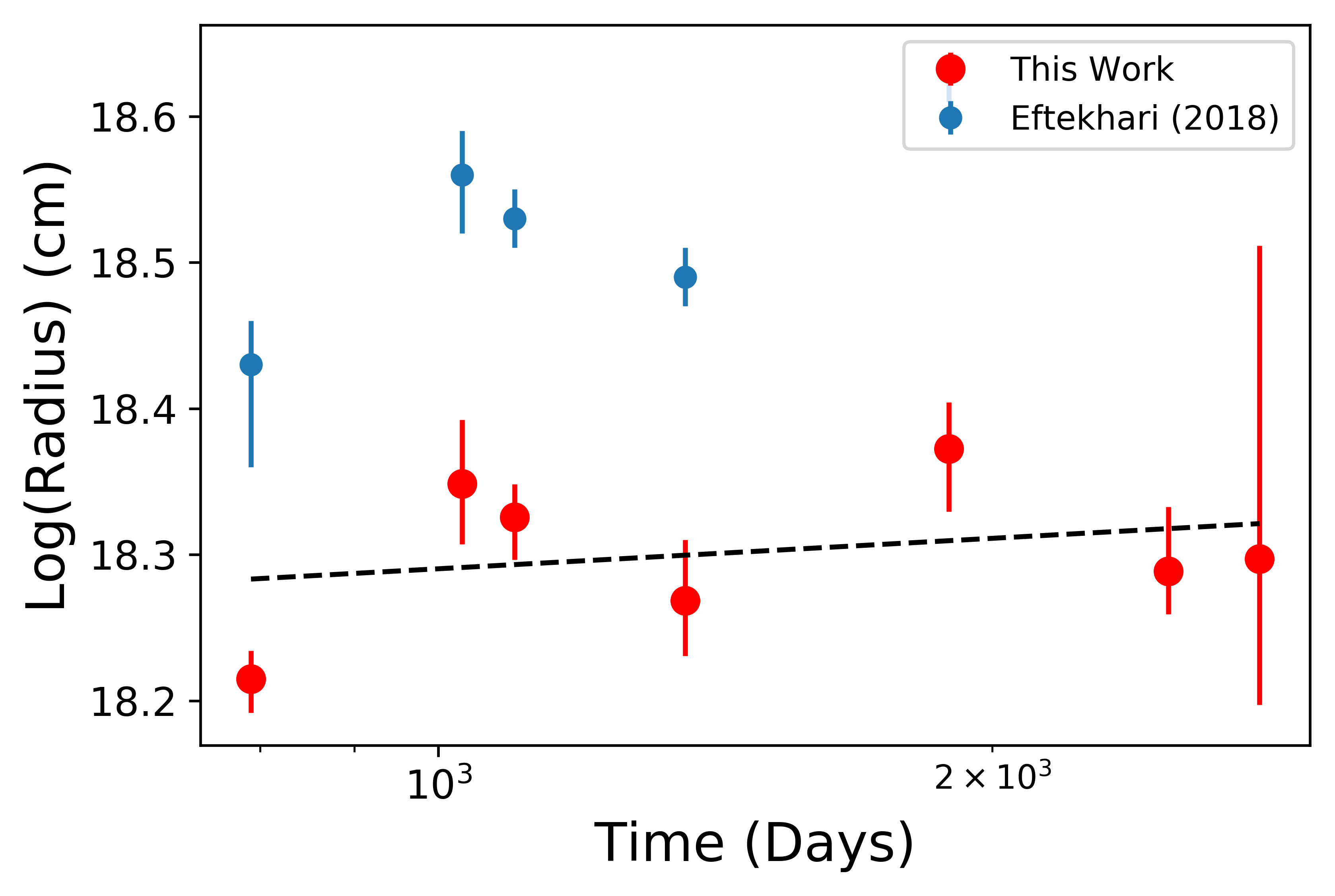}
    \label{fig:rad}
    \includegraphics[width=.75\columnwidth]{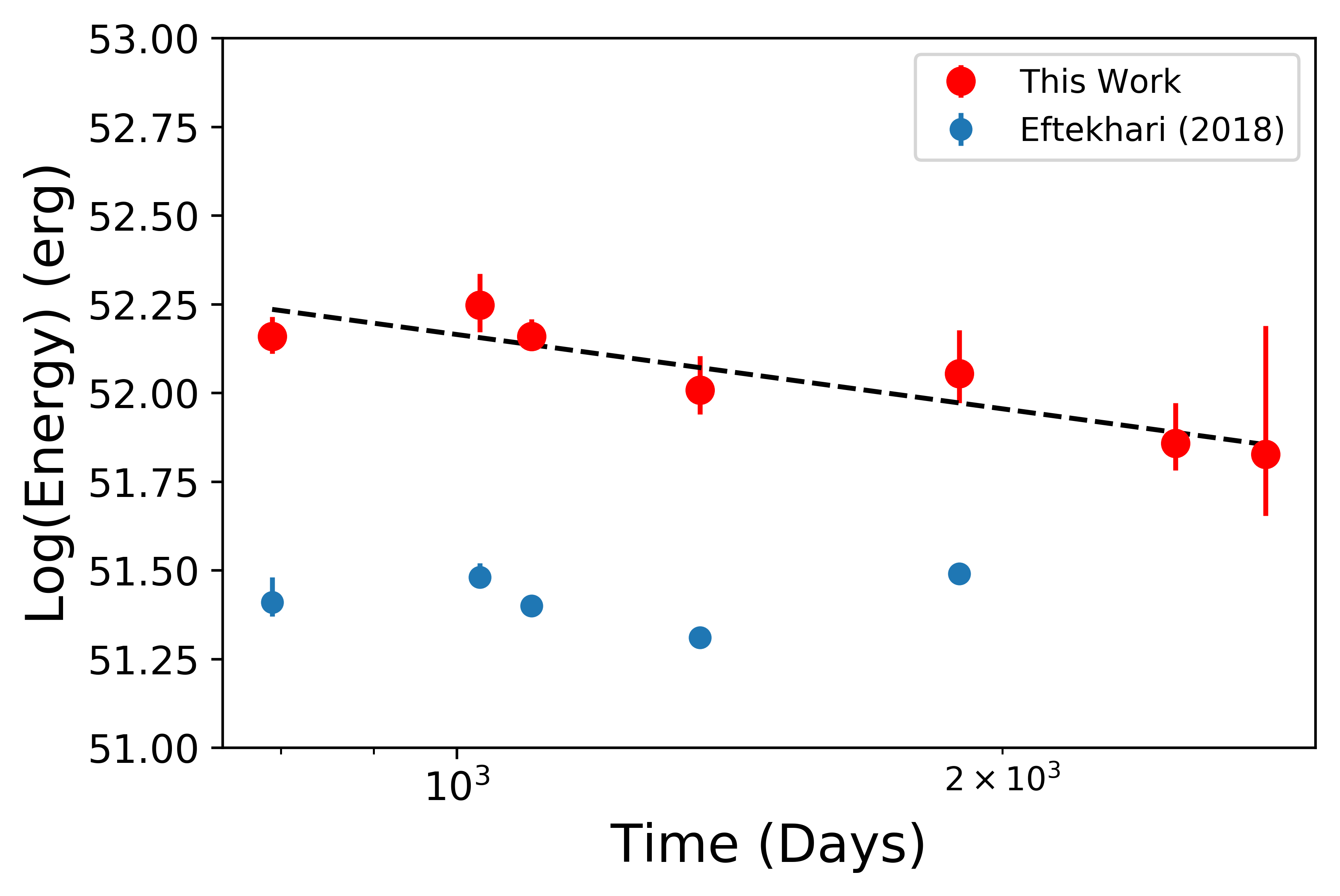}
    \label{fig:energy}
    \includegraphics[width=.75\columnwidth]{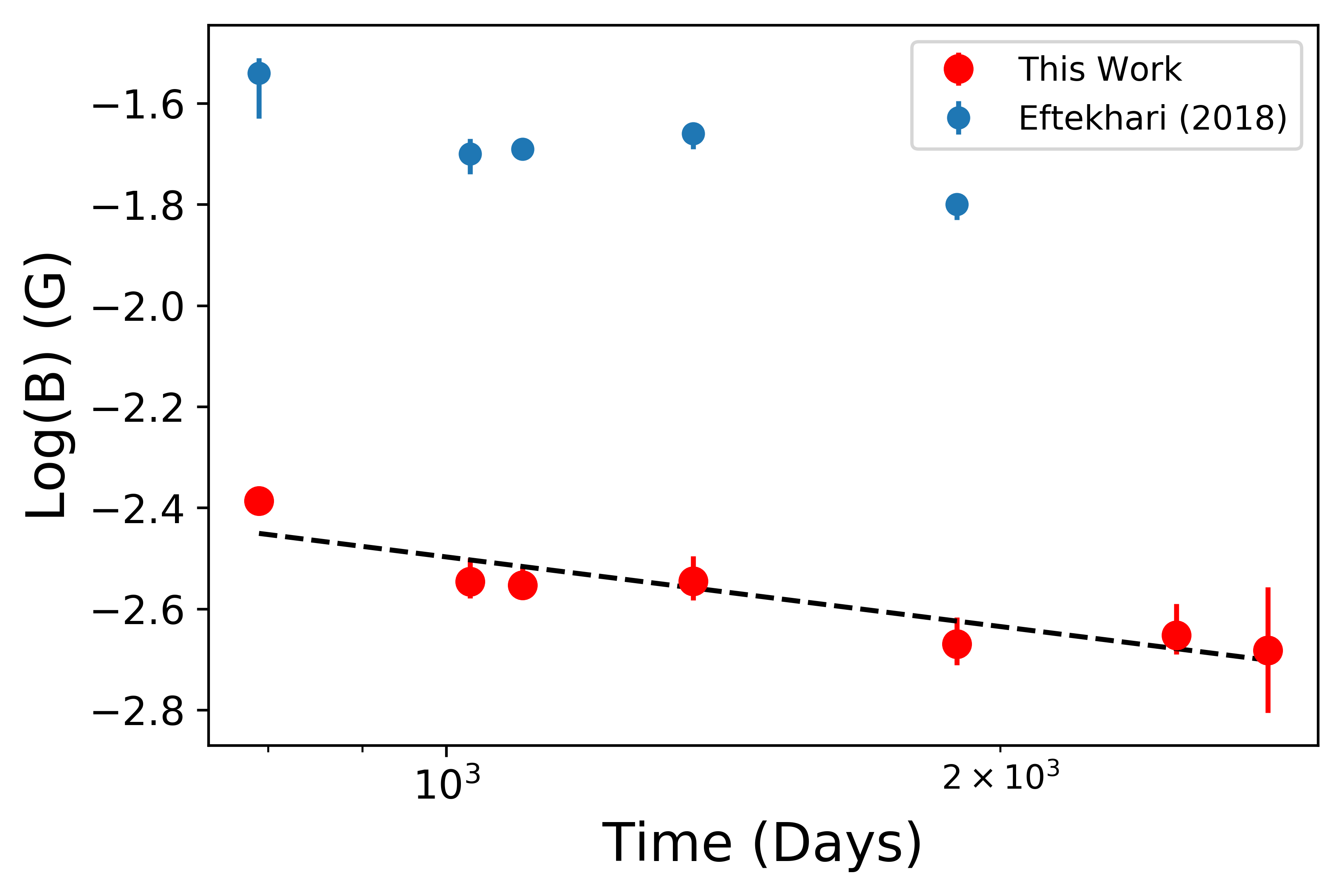}
    \label{fig:mag}
    \includegraphics[width=.75\columnwidth]{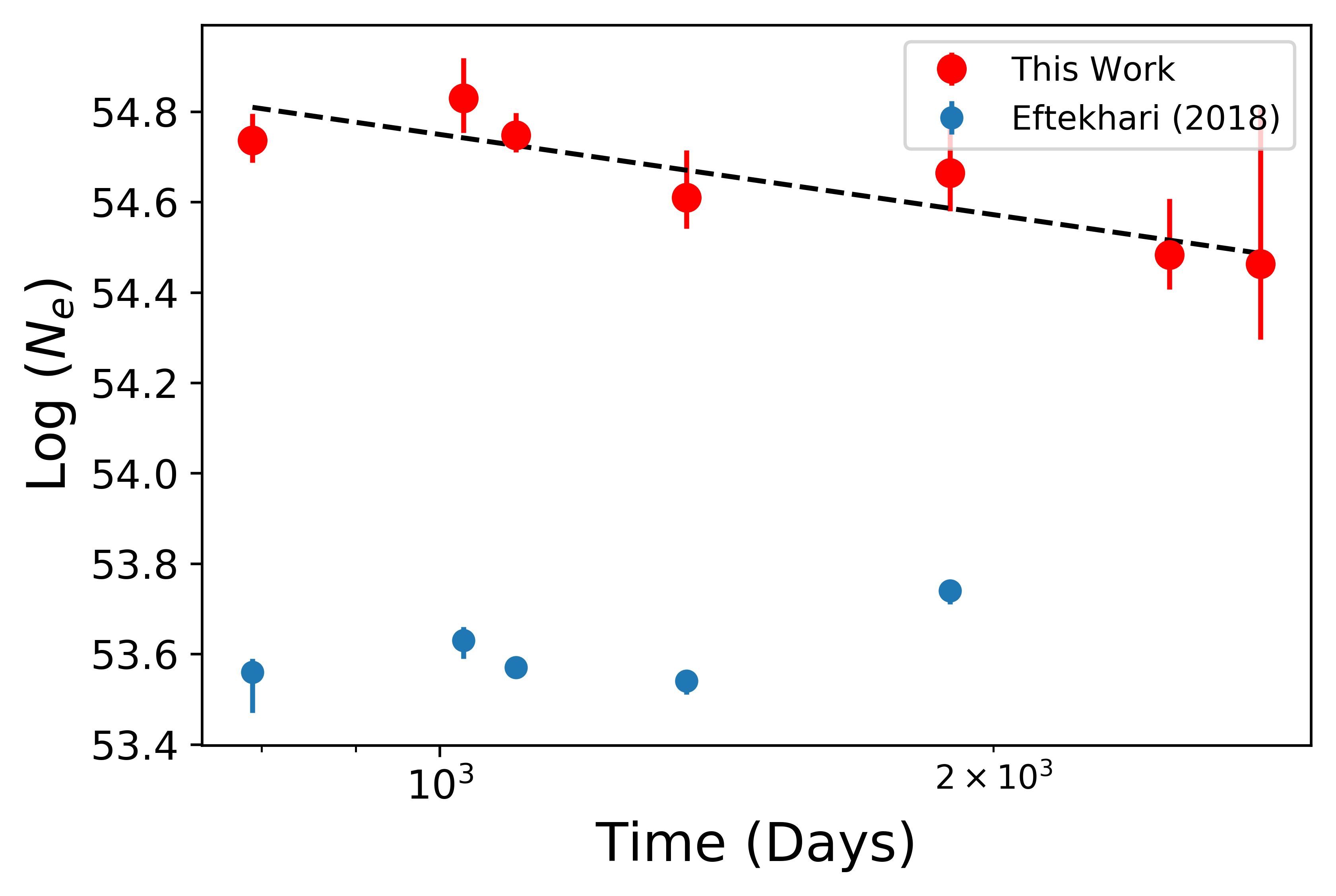}
    \label{fig:Ne}
    \includegraphics[width=.75\columnwidth]{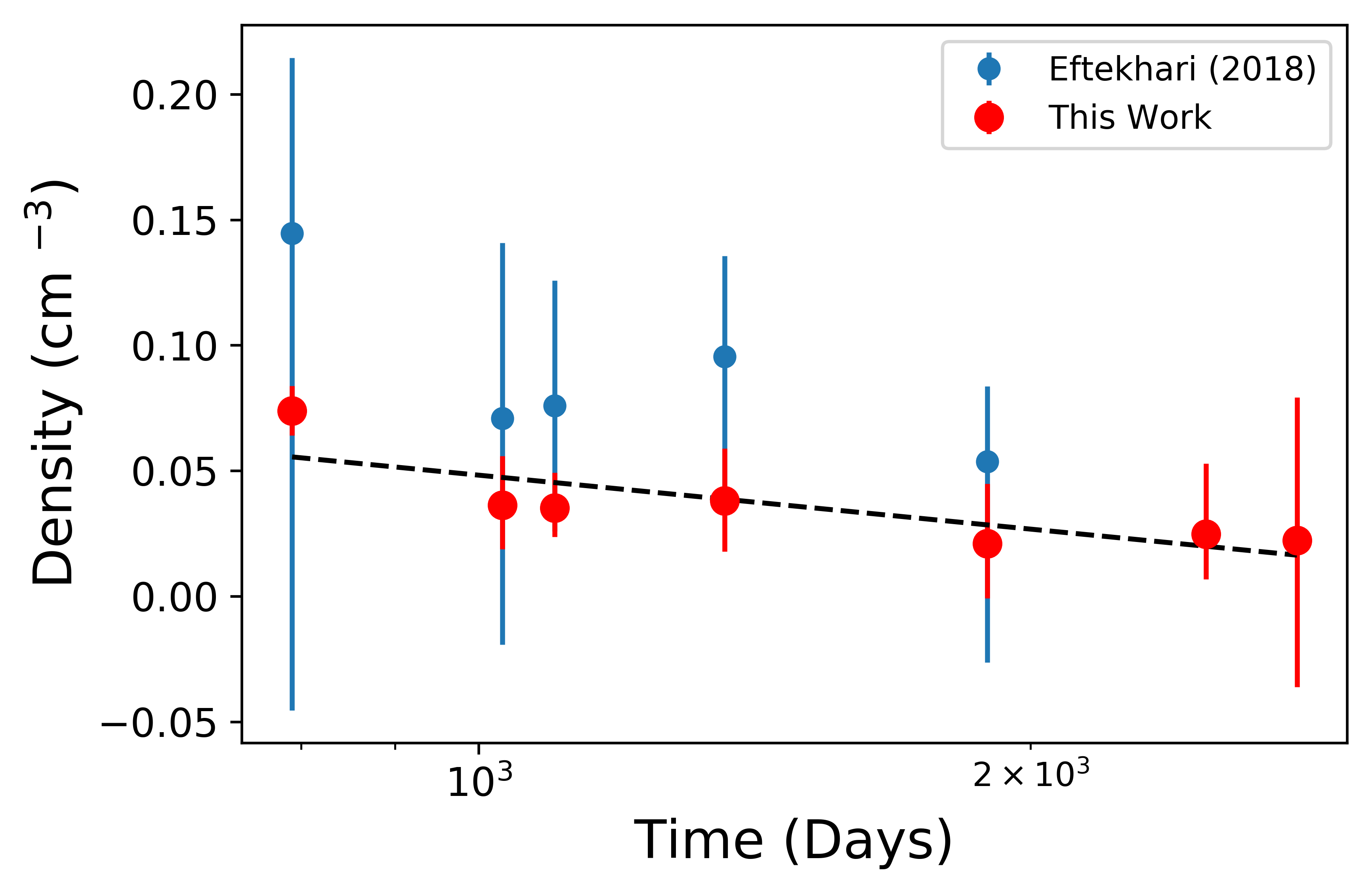}
    \label{fig:dens}
    \caption{The radius, kinetic energy, magnetic field, number of radiating electrons, and ambient density from our equipartition analysis (red).  We also show the results for the same parameters from  \citet[][blue]{p3}.  The two sets of values exhibit the same temporal trend, but differ in normalization due to the use of $p=3$ in this work compared to $p=2.5$ in \citet{p3}.}
    \end{center}    
\end{figure}

The inferred outflow and external medium parameters as a function of time are plotted in Figure~\ref{fig:dens}, using $\epsilon_B = 0.01$ (see \S\ref{sec:cooling}); we include the values reported in \citet{p3} using $p=2.5$ for reference.  For the radius, we find values that are about a factor of 1.5 times smaller than in \citet{p3}, but the trend of $R\propto t^{0.1}$, which has decreased when compared to the $R\propto t^{0.2}$ found by \citet{p3}. For the energy, our values are about 5 times larger than in \citep{p3}, due to both the changes $p$ and $\epsilon_{B}$. We find a steady decline in the energy with $E\propto t^{-0.7}$.  The magnetic field strength is about 6 times lower than in \citet{p3}, and exhibits a steady decline with $B\propto t^{-0.5}$. Finally, we find values of $n_{e,\rm ext}$ are lower than in \citet{p3} by a factor of 2, although within the error bars for that work, and declining with $n_{e,\rm ext} \propto t^{-0.07}$.

\subsection{Cooling Frequency and $\epsilon_{B}$}
\label{sec:cooling}

\citet{p3} reported a deviation from equipartition in  by searching for a cooling break using the optical/NIR and late-time ($\delta t\gtrsim 500$ d) X-ray data.  They find $\epsilon_B\approx 10^{-3}$, two orders of magnitude lower than expected for equipartition (relative to the assumed value of $\epsilon_e=0.1$). However, because here we use $p=3$, thereby changing the steepness of the SED slope above $\nu_p$, we expect that this will lead to a change in the location of the cooling frequency, $\nu_{c}$, and hence $\epsilon_B$.  The cooling frequency is given by \citep{Sari1998}:
\begin{equation} 
\nu_c = 2.8\times10^{6} \gamma_c^2 \Gamma^2 B,
\end{equation}
where $\gamma_c = 5\times10^9 \epsilon_B^{-1} t_d^{-1} n_{\rm ext}^{-1}$.  

\begin{figure}
    \includegraphics[width=\columnwidth]{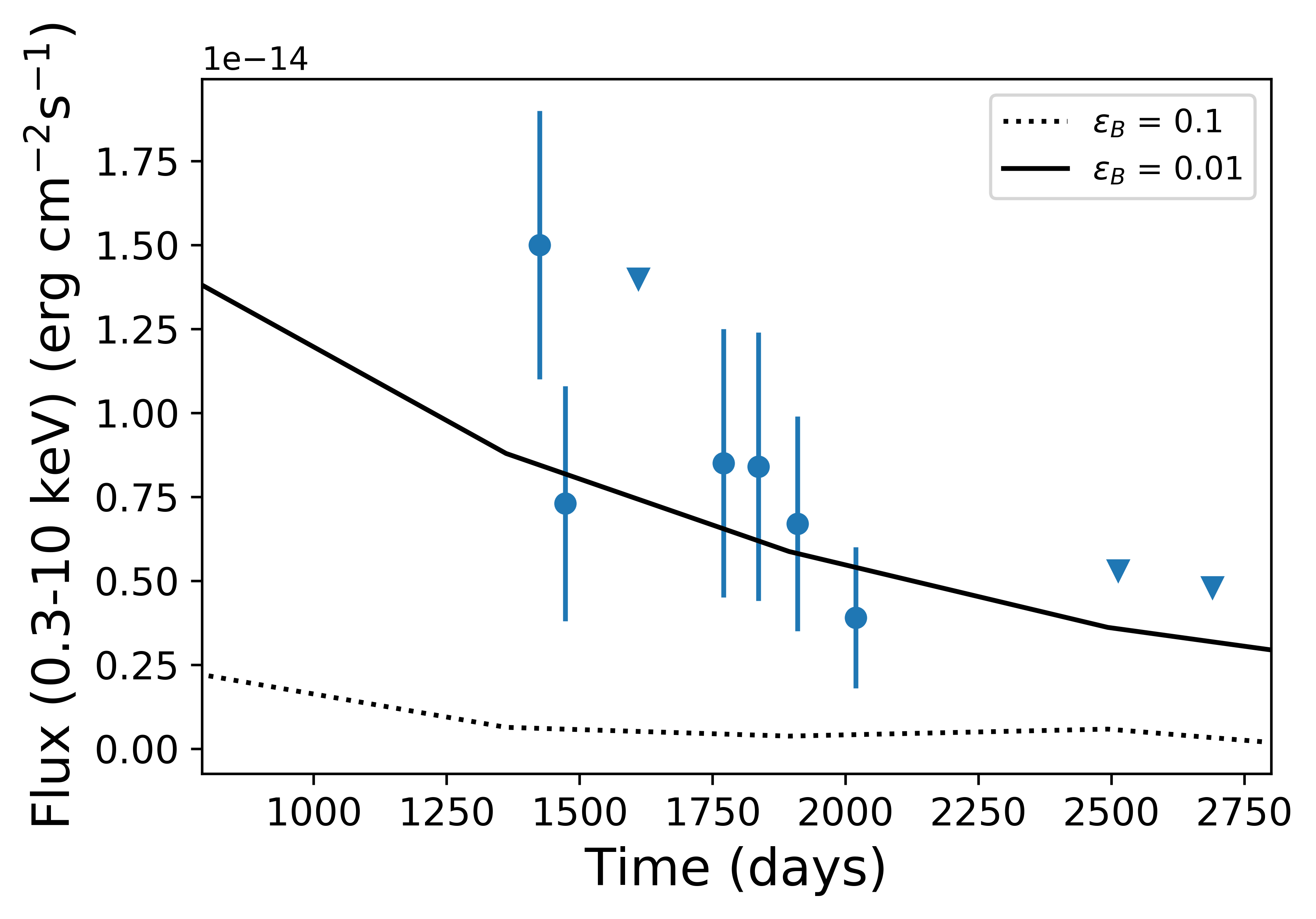}
    \caption{Late-time X-ray light curve, including the two upper limits measured here (blue).  The lines are an extrapolation of the radio SEDs to the X-ray regime, assuming equipartition ($\epsilon_{B}=0.1$; black dotted line) and a deviation from equipartition, with $\epsilon_{B}=0.01$ (solid black line). The latter matches the X-ray observations.}
    \label{fig:cooling}
\end{figure}

In Figure~\ref{fig:cooling} we show the X-ray data at $\gtrsim 1400$ d along with an extrapolation of the radio models assuming equipartition ($\epsilon_B=0.1$).  We find that this model underpredicts the observed X-ray flux.  Instead, a model with $\epsilon_{B}\approx 0.01$ provides an adequate fit to both the amplitude and time evolution of the X-ray flux; see Figure \ref{fig:cooling}.  The blastwave and external medium parameters provided in the previous sections adopt this departure from equipartition.

\section{Discussion}
\label{sec:discussion}

\sw\ continues to be the best and longest studied relativistic TDE to date.  Below we summarize the on-going evolution of the outflow and the ambient density profile, briefly discuss the implications of the VLASS detections, and cover what we can expect from observations of \sw\ in the future.

\subsection{The Ongoing Evolution of \sw}
\label{sec:evolution}

After Day $\sim 700$ the outflow from \sw\ transitioned to non-relativistic expansion \citep{p3}.  Our most recent observations confirm this trend.  The steady energy scale of the outflow on a timescale of $\sim 800-2800$ days (Figure~\ref{fig:energy}) indicate no significant injection of energy from initially slow ejecta that may have caught up with the blastwave at late times.  Similarly, the continued decline in X-ray emission (with the non-detections presented here) indicates that this band is still dominated by forward shock emission, and a jet has not been re-activated.


\begin{figure}
    \includegraphics[width=1.05\columnwidth]{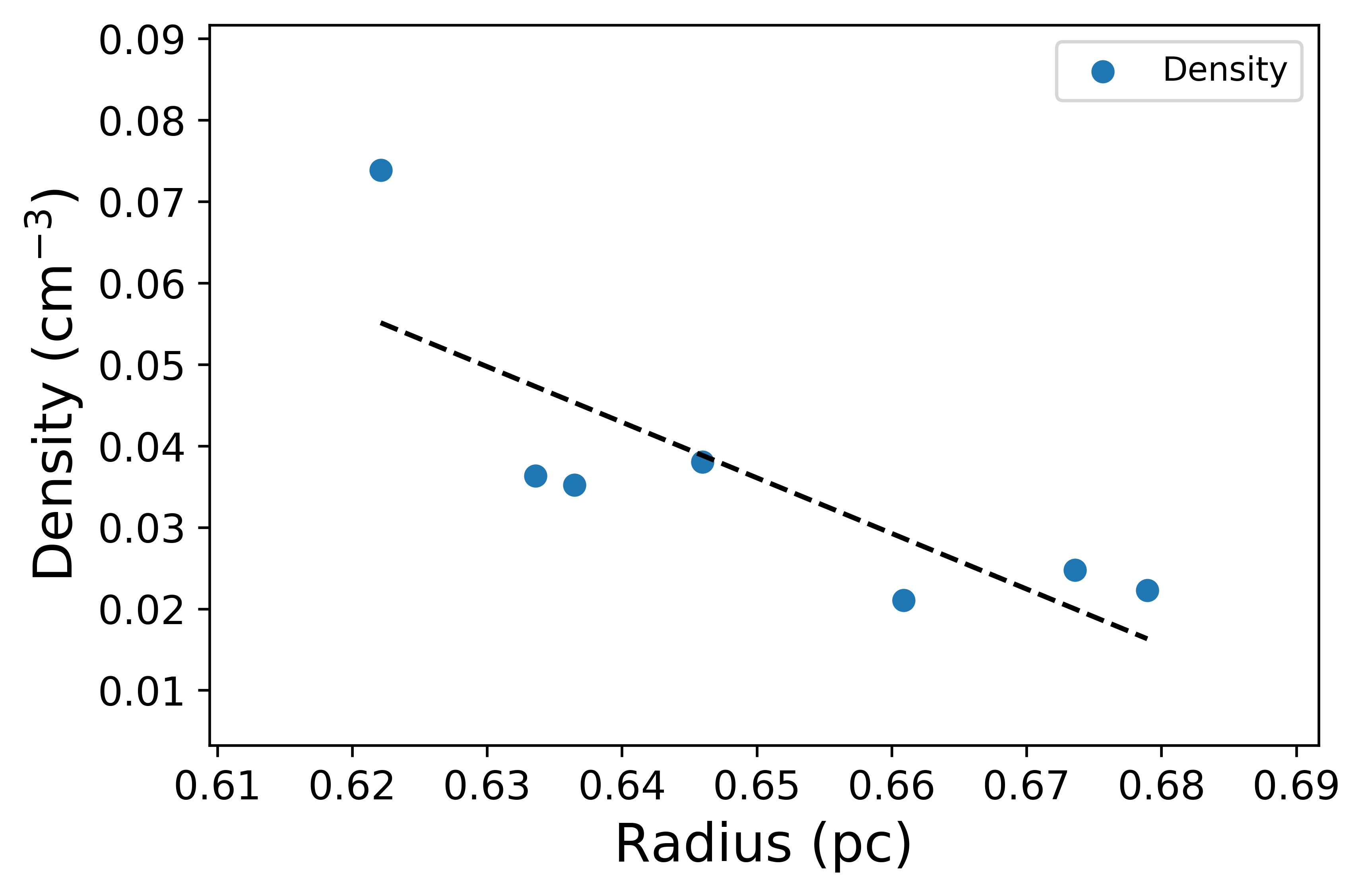}
    \label{fig:dens-sw-only}
    \includegraphics[width=1.05\columnwidth]{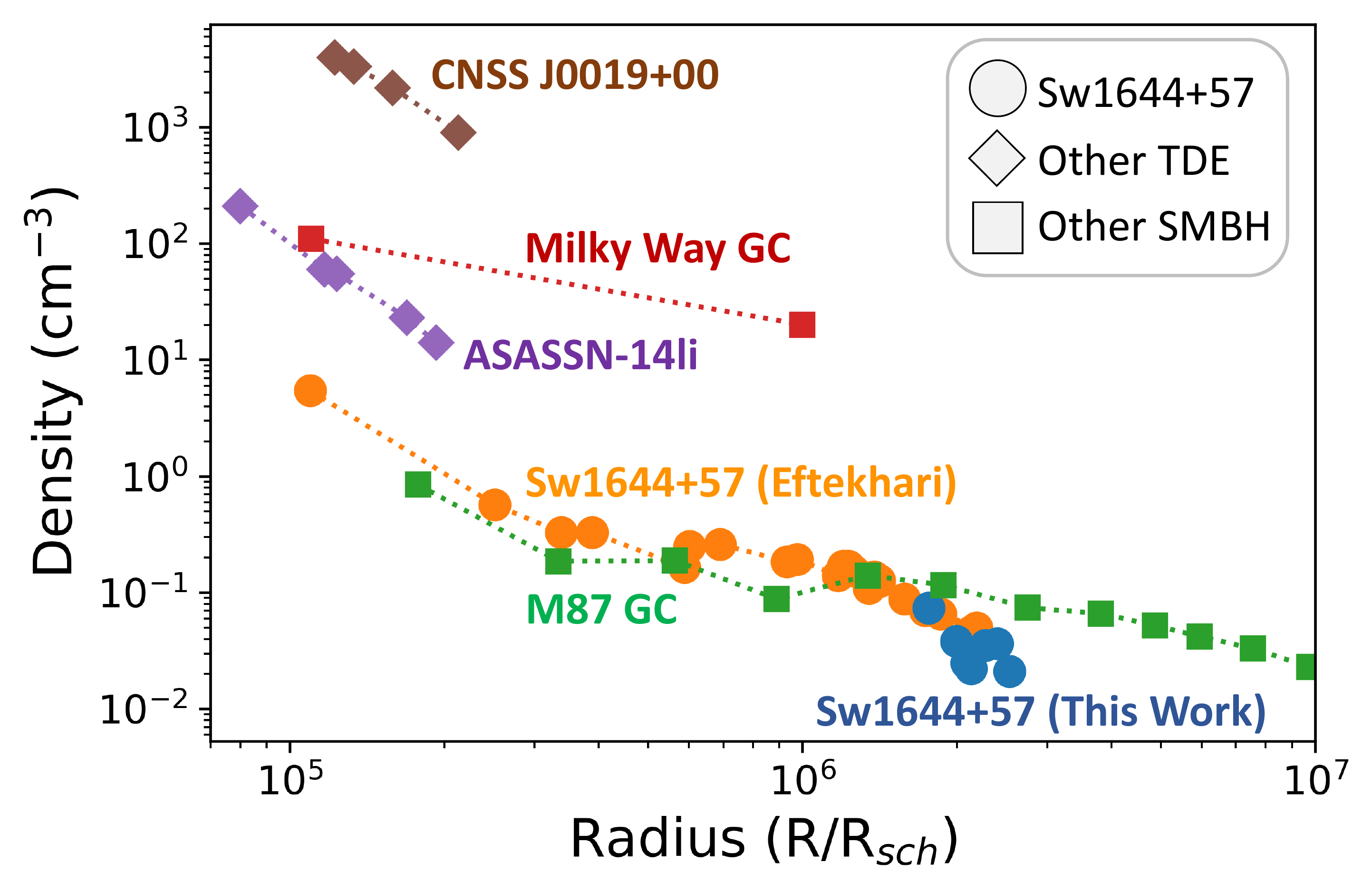}
    \label{fig:dens-rad}
    \caption{The circumnuclear density profile as a function of radius in parsecs ({\it Top}) and in units of $R_{\rm sch}$ ({\it Bottom}; using $M_{\rm SMBH}=10^{6.5}$ M$_\odot$).  The data at smaller radii (from \citealt{p3}; yellow) are scaled by the difference in radius and density calculated in \ref{sec:modeling}.  For reference, we also include data for Sgr A* \citep[red;][]{Quataert2004}, M87 \citep[green;][]{Russell2015}, and two TDEs with non-relativistic outflows (ASASSN-14li: \citealt[][purple]{Alexander2016}; CNSS J0019+00: \citealt[][brown]{Anderson2019}).}
\end{figure}

The continued radio observations extend our ability to measure the density profile around the supermassive black hole to a larger radius of about 0.7 pc.  Using the time evolution of the blastwave radius and density of emitting particles, we find that the density profile on a scale of $\approx 0.5-0.7$ pc roughly follows $n_{ext} \propto r^{-2}$; see Figure~\ref{fig:dens-rad} (top).  For the purpose of comparison with other supermassive black holes, we also scale the density profile to the Schwarzschild radius of the black hole (Figure~\ref{fig:dens-rad}; bottom).  We find that the density in the environment of \sw\ is about 100 times lower than around Sgr A*, but comparable to that of M87.  For reference, we also include two TDEs with non-relativistic outflows, ASASSN-14li \citep{Alexander2016} and CNSS J0019+00 \citep{Anderson2019}, which overlap with the distances probed by \sw\ at early times.  We find that both of those events occurred in higher density regions than \sw.

\subsection{VLASS Detection and Implications for Blind TDE Searches}
\label{sec:vlass}

\begin{figure}
    \includegraphics[width=\columnwidth]{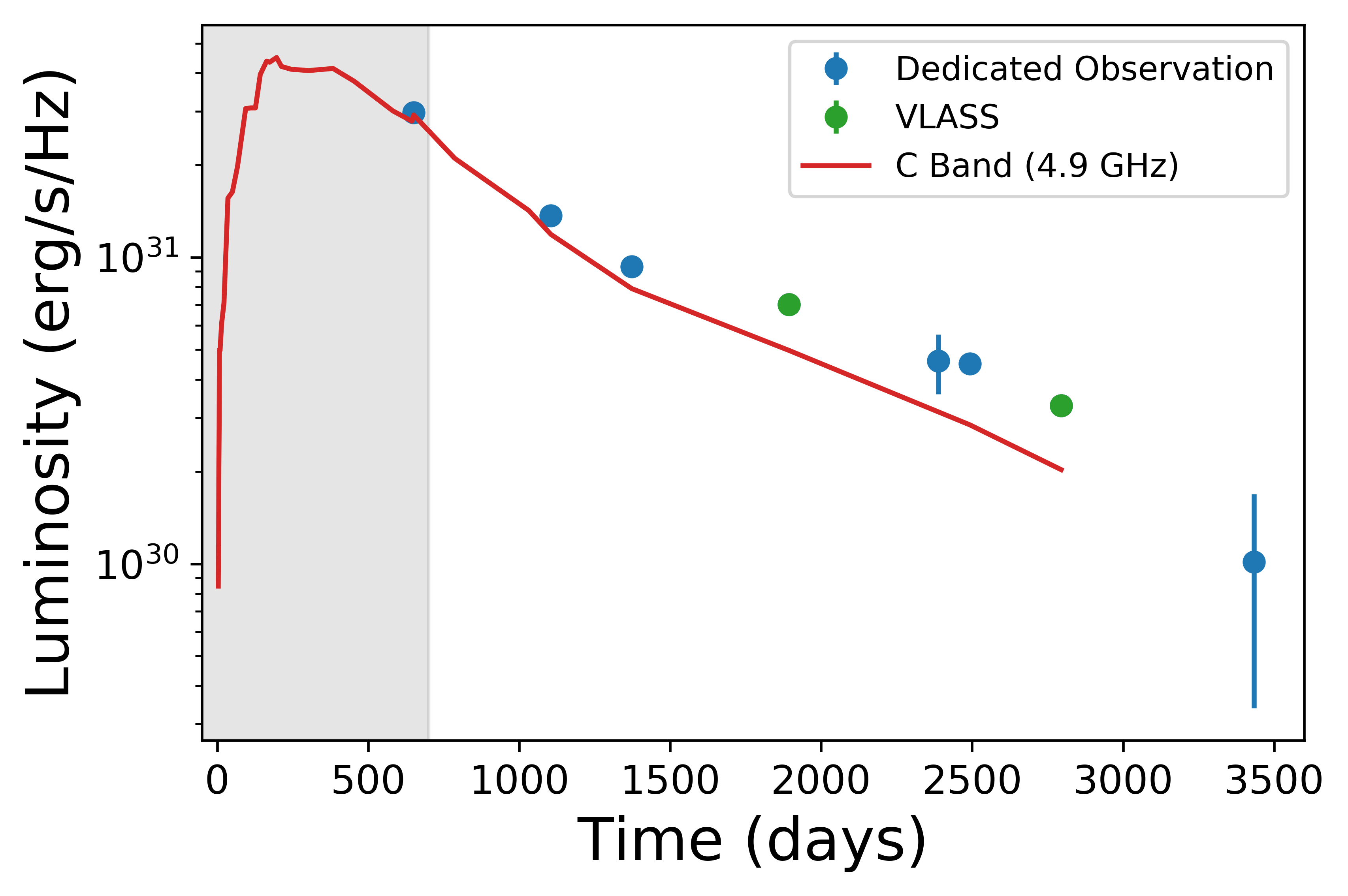}
    \caption{The radio spectral luminosity in S-band ($\sim$3 GHz; data points), and at C-band (4.9 GHz; red; \citealt{p1,p2,p3}).  The shaded region indicates the time when the outflow is relativistic \citep{p3}.}
    \label{fig:S-band-lumin}
\end{figure}

The long-term radio emission from \sw\ raises the possibility of finding similar relativistic TDEs through their radio emission, even if the jet is initially oriented off-axis.  \citet{Metzger2015} discussed the possibility of finding on-axis TDEs like \sw\ in the VLASS survey and find $\sim 1$ detectable event assuming a light curve duration of $\sim 10^3$ days and a typical off-axis peak luminosity about an order of magnitude lower than for \sw.  Recently, \citet{Anderson2019} reported the discovery of a radio transient (CNSS J001947.3+003527) in Stripe 82 survey data, which they interpret as an off-axis TDE.  This event had a peak luminosity of $5\times 10^{28}$ erg s$^{-1}$ Hz$^{-1}$, about three orders of magnitude lower than \sw.

Our two VLASS detections of \sw, combined with our dedicated observations provide unique insight in the context of TDE blind radio detections.  In Figure~\ref{fig:S-band-lumin} we shows the light curve of \sw\ in S-band (3 GHz) and C-band (5 GHz). The spectral luminosity range traced by our observations so far is $\approx 3\times 10^{31}$ erg s$^{-1}$ Hz$^{-1}$ at a peak time of $\sim 0.55$ yr to $\approx 10^{30}$ erg s$^{-1}$ Hz$^{-1}$ at about 9.5 yr.  The outflow from \sw\ became non-relativistic on a timescale of $\approx 2$ yr \citep{p2,p3}, which can be used as a characteristic timescale for when similar events would become detectable even for an initial off-axis orientation; the spectral luminosity on this timescale was $\approx 2\times 10^{31}$ erg s$^{-1}$ Hz$^{-1}$.

Given the slow cadence of a survey like VLASS ($\approx 2.5$ yr between successive epochs), it is most likely that similar events to \sw\ will be captured on a timescale of $\sim{\rm yr}$ post-disruption.  Thus, an event with a similar luminosity will be detectable at $10\sigma$ ($\approx 2$ mJy) to a redshift of $z\approx 0.6$ for an on-axis event.  The radio emission from \sw\ post-peak evolves roughly as $t^{-1}$, so over the timescale of 2 VLASS epochs we expect that a similar source will fade by a factor of $\sim 2-3$ (depending on the phase of the light curve captured in the first epoch).  At an initial detection of $10\sigma$, this would be a measurable fading.  Thus, we conclude that events like \sw\ can be captured in VLASS and shown to be fading to $z\sim 0.6$.  Using the same approach for off-axis events, the lower luminosity means these events will be found in VLASS at $z\lesssim 0.5$.

Updating this distance with the calculation in \citet{Metzger2015}, we find that there should be $\sim 2$ jetted TDE visible in VLASS.  In the case of off-axis TDEs, the difference in rate would be $1.5\times$ lower due to the difference in comoving volumes between $z\sim 0.5$ and $z\sim 0.6$.  Assuming a beaming factor of $\sim 10^2$, this yields $\sim 10^2$ off-axis TDE events in the VLASS survey.  However, we emphasize that our estimate assumes the evolution of \sw\ is generic, and there may be a more significant difference between peak luminosity for on-axis and off-axis TDEs.

\subsection{Future Observations}

\sw\ is continuing to evolve as a non-relativistic outflow with an energy of $\approx 10^{52}$ erg into a ambient medium with a density of $\approx 0.1$ cm$^{-3}$ on a scale of about 0.65 pc from the black hole (Figure \ref{fig:dens-sw-only}).  As we noted previously (e.g., \citealt{p3}), the radio emission will continue to be detectable with the VLA for decades.  With a flux density of about 1 mJy at 3 GHz and a decline rate of about $t^{-1}$, the emission should be detectable for an order of magnitude longer in time (i.e., $\sim {\rm century}$).

In terms of continued characterization of the outflow, it is essential to capture the peak of the SED, which is evolving to lower frequencies.  In our latest multi-frequency observation at 7.7 yr post-disruption, we find that $\nu_p\approx 1.3$ GHz, which is at the lower frequency end of the VLA's sensitive bands.  We expect \sw\ to be detectable at $5\sigma\sim 1$ mJy at P-band (0.35 GHz) in a few hours of dedicated observing with the VLA or $\sim 10$ hr of commensal observing with VLITE.  Another facility with sufficient sensitivity below 1 GHz is the Giant Meterwave Radio Telescope (GMRT), and we will report such follow-up observations in the future.  At even lower frequencies, the Low-Frequency Array (LOFAR) is not sensitive enough to detect $\sim {\rm mJy}$ level emission, and the next generation VLA (ngVLA) proposal does not cover sub-GHz frequencies.  We note that \sw\ is not visible from the future Square Kilometer Array (SKA) sites.  As such, while the radio emission at GHz frequencies will continue to be detectable for decades, it is possible that a detailed characterization of the outflow may not be possible. Still, tracking the optically thin portion of the SED will allow us to determine changes in the dynamical evolution of \sw, or evolution in the electron power law index, $p$.  

Finally, the proposed Lynx X-ray Observatory may be able to detect emission from \sw, depending on when (and if) it is launched, as well as whether \sw\ continues to evolve in X-rays at its present rate.

\section{Conclusions}
\label{sec:conclusions}


We presented new VLA radio data ($1.5-22$ GHz) and \textit{Chandra} X-ray data extending to about 3400 days post-disruption, and used these observations to determine the continued dynamical evolution of the outflow and the density of the ambient medium.  The combination of continued radio detections and X-ray upper limits allow us to characterize the synchrotron spectral energy distribution. We find that, unlike at early time, the electron power law at $\gtrsim 10^3$ days index is $p\approx 3$.  The outflow continues to exhibit non-relativistic expansion, with a radius of about $2\times 10^{18}$ cm (0.65 pc) and a kinetic energy of about $10^{52}$ erg.  The circumnuclear density is about $0.1$ cm$^{-3}$.

We also report two detections of \sw\ in VLASS data, making it the first TDE detection in this sky survey (albeit not a blind detection).  We estimate, based on the long-term evolution of \sw, that for events with similar luminosities and durations, $\sim 2$ on-axis and $\sim 10^2$ off-axis TDEs may be present in the VLASS data.  We note, however, that \sw\ remains by far the most luminous TDE at radio frequencies so the detection rate in VLASS may be much lower.  In addition, even if $\sim 10^2$ TDEs are present in the VLASS data, identifying these sources within the large survey footprint will not be trivial.

As \sw\ continues its slow fade, it will detectable for decades with the VLA (and its successors), although the peak frequency continues to evolve to sub-GHz frequencies that are challenging to measure at the current brightness of \sw.  Still, continued observations of this unique event are essential to determine the long-term evolution of the outflow and the structure of the circumnuclear medium on sub-pc scales.

\section{Acknowledgements}
We thank Kate Alexander and Griffin Hosseinzadeh for conversations and input regarding the manuscript, and Wendy Lane for assistance with VLITE data. Support for this work was provided by the National Aeronautics and Space Administration through Chandra Award Numbers GO8-19087X and GO8-19098X issued by the Chandra X-ray Center, which is operated by the Smithsonian Astrophysical Observatory for and on behalf of the National Aeronautics Space Administration under contract NAS8-03060. The National Radio Astronomy Observatory is a facility of the National Science Foundation operated under cooperative agreement by Associated Universities, Inc.  The scientific results reported in this article are based in part on observations made by the Chandra X-ray Observatory.

\bibliography{bibliography}
\end{document}